\documentclass[twocolumn,prd,showpacs]
              {revtex4}

\usepackage{graphicx}

\begin{document}

\title{Particle escape into extra space}

\author{Michael~Maziashvili}

\email{maziashvili@ictsu.tsu.edu.ge} \affiliation{Department of
Theoretical Physics, Tbilisi State University,\\ 3 Chavchavadze
Ave., Tbilisi 0128, Georgia \\Institute of High Energy Physics and
Informatization,\\ 9 University Str., Tbilisi 0186, Georgia }

\begin{abstract}
We focus on escape of a spin integer particle the challenge for
which is of course that the corresponding field equation contains
the second order time derivative and, in general, may be
problematic for interpreting the extra-dimensional part of the
field as a wave function for the KK modes as it is usually
regarded.
\end{abstract}


%11.10.Kk. Field theories in dimensions other than four.

\maketitle

Extra-dimensional models provide a fertile approach to a broad
class of problems in the theory of elementary particles and
cosmology. In the brane-wirld model building the question of
paramount importance is the localization of the Standard Model
particles on the brane. We continue our attempt to put forward the
theoretical framework for describing the dynamics of particle
escape into extra space. The troublesome aspect of our approach
presented in \cite{Ma} has to do with the time evolution of
extra-dimensional part of the scalar field. Namely, it does not
preserve the norm of the metastable state and therefore
contradicts the wave function interpretation. Let us strive to
maintain this interpretation of exra-dimensional part of the field
as underpinning for the brane-world model and try to describe the
time evolution of the quasy-localized state.

To be more concrete in what follows let us focus on the
brane-world model with non-factorizable warped geometry
\cite{GoRa}
\begin{equation}\label{GRS}ds^2=e^{-2\kappa|z|}\eta_{\mu\nu}dx^{\mu}dx^{\nu}-dz^2~,\end{equation}
where the parameter $k$ is determined by the bulk cosmological and
five-dimensional gravitational constants respectively. In making
the transition from the bulk with coordinates $(x^{\mu},~z)$ to
the brane with local coordinates $x^{\mu}$, the field
$\phi(x^{\mu},~z)$ is usually decomposed into four end
extra-dimensional parts
\begin{equation}\label{freep}\phi(x^{\mu},~z) \sim e^{ip_{\mu}x^{\mu}}\psi_m(z)~,\end{equation} where $\psi_m(z)$ is
understood as a wave function responsible for localization of the
KK modes \cite{BG} with the mass $p_{\mu}p^{\mu}=m^2$. The
transverse equation governing $\psi(z)$ has the form of stationary
Schr\"{o}dinger equation. The scalar product in space of $\psi
(z)$ functions arises naturally after substituting of
Eq.(\ref{freep}) into the action functional and separating the
integration with respect to the extra and four-dimensional
variables
\[\left<g|f\right>=\int\limits_{-\infty}^{\infty}dze^{-2k|z|}g^*(z)f(z)~.\]
The brane Eq.(\ref{GRS}) localizes the massless scalar field as it
was shown in \cite{BG}, but when this field is given bulk mass
term the bound state becomes metastable against tunneling into an
extra dimension \cite{DRT}. Let us presume that the metastable
states are defined by the Gamow's method as it was done in
\cite{DRT}. namely, if the transverse equation admits tunneling of
initially brane localized particle into extra space one finds the
complex eigenvalues in the mass spectrum by imposing the outgoing
wave boundary condition. To gain a complete picture it should be
noticed that for the Gamow's method the concrete form of the time
evolution of the state encoded in the Schr\"{o}dinger equation is
a crucial moment. So following the Gamow's method it seems
reasonable to take the following \emph{ans\"{a}tze} as a
satisfactory point of departure
\begin{equation}\label{plwdec}\phi = e^{i(E_0t-\vec{p}\,\vec{x})}g(t,~z)~,~\mbox{with}~~g(t,~z)=e^{-iEt}\psi(z)~,\end{equation} where $\vec{p}$
is the three momentum of the particle,
$E_0=\sqrt{m_0^2+\vec{p}\,^2}$ is its energy and $m_0$ is defined
through the Gamov's method (for simplicity only one metastable
mode $m_0$ is assumed). The four-dimensional plane wave in
Eq.(\ref{plwdec}) stands for a free particle with the mass
$(p^{\mu} p_{\mu})^{1/2}$ moving in $x^{\mu}$ space along
$\vec{p}$ while $g(t,~z)$ describes its localization properties
across the brane which, in general, may vary in course of time.
For this \emph{ans\"{a}tze} one gets the following equation
\begin{equation}\label{geigeq}-\partial^2_z\psi+4k\,\mbox{sgn}(z)\partial_z\psi+\mu^2\psi=
e^{2k|z|}\left((E-E_0)^2-p^2\right)\psi~,\end{equation} where
$\mu$ is the bulk mass parameter \cite{DRT}. Following the paper
\cite{DRT} under assumption $\mu \ll k$ one finds without much ado
\[E=2E_0-{im_0\Gamma\over \sqrt{\vec{p}\,^2+m_0^2}}~,~~{\Gamma\over m_0}={\pi\over
8}\left({m_0\over k}\right)^2~,~~m_0={\mu\over \sqrt{2}}~,\] and
the corresponding wave function satisfying the outgoing wave
boundary condition. In this way one finds that the probability of
finding the particle on the brane decays exponentially in time and
the corresponding lifetime is given by
$\sim\sqrt{\vec{p}\,^2+m_0^2}/m_0\Gamma$.

But, as it is well known the precise quantum mechanical
consideration of metastable state decay based on the approach
proposed by Fock and Krylov \cite{FK} leads to the deviation from
exponential decay law for small and large values of time. More
precisely in quantum mechanics it is well established that an
exponential decay cannot last forever if the Hamiltonian is
bounded below and cannot occur for small times if, besides that,
the energy expectation value of the initial state is finite
\cite{CSMFGR}. Let us consider this problem for the brane-world
model following to the general \emph{ans\"{a}tze}
\[\phi = e^{i(E_0t-\vec{p}\,\vec{x})}g(t,~z)~.\]
The equation of motion for
\[f(t,~z)= e^{iE_0t}g(t,~z)\] takes the form
\begin{equation}\label{gtimequ}(\partial^2_t+p^2)f-e^{2k|z|}\partial_z(e^{-4k|z|}\partial_zf)+e^{-2k|z|}\mu^2f=0~,\end{equation}
the general solution for which is given by \cite{Lee}
\begin{eqnarray}\label{gensol}f(t,~z)&=&\int\limits_{-\infty}\limits^{\infty}G_1(t,~z,~z')f(0,~z')dz'\nonumber\\
&+&\int\limits_{-\infty}\limits^{\infty}G_2(t,~z,~z')\dot{f}(0,~z')dz'~,\end{eqnarray}
where
\begin{eqnarray}G_1(t,~z,~z')=e^{-2k|z'|}\int\limits_p\limits^{\infty}dE\cos(Et)\varphi_p(E,~z)\varphi_p(E,~z')~,\nonumber\\
G_2(t,~z,~z')=e^{-2k|z'|}\int\limits_p\limits^{\infty}dE{\sin(Et)\over
E}\varphi_p(E,~z)\varphi_p(E,~z')~,\nonumber\end{eqnarray} and the
functions
\[\varphi_p(E,~z)=\sqrt{{E\over m}}\,\varphi(m,~z)~,\]
with $m=(E^2-p^2)^{1/2}$ are determined by the normalized
eigenfunctions of
\begin{equation}\label{eigeq}-\partial^2_z\varphi+4k\,\mbox{sgn}(z)\partial_z\varphi+\mu^2\varphi=
e^{2k|z|}m^2\varphi~,\end{equation} satisfying the junction
condition across the brane
\[\varphi=\sqrt{{m\over 2k}}e^{2k|z|}\left[a(m)J_{\nu}\left({m\over k}e^{k|z|}\right)+b(m)Y_{\nu}\left({m\over k}e^{k|z|}\right)\right]~,\]
with the index $\nu=\sqrt{4+\mu^2/k^2}$ and coefficients
\[a(m)=-{A(m)\over \sqrt{1+A^2(m)}}~,~~b(m)={1\over \sqrt{1+A^2(m)}}~,\]

\[A(m)={Y_{\nu-1}(m/k)-(\nu-2)(k/m)Y_{\nu}(m/k)\over J_{\nu-1}(m/k)-(\nu-2)(k/m)J_{\nu}(m/k)}~.\]

To compute $f(t,~z)$ one needs to know the initial data
$f(0,~z),~\dot{f}(0,~z)$. In general, arbitrarily taking the
initial velocity $\dot{f}(0,~z)$ we face the problem that the norm
of $f(t,~z)$ may not be preserved \cite{Ma}.

\begin{figure}[t]
\includegraphics[width=\columnwidth]{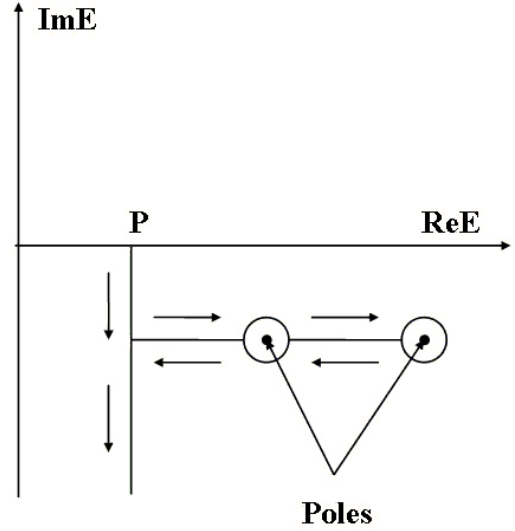}\\
\caption[lifetime]{The integration contour.}
\end{figure}

With the above comment in mind, to keep as close as possible to
the time dependence of Schr\"{o}dinger equation, the further
insight into the evaluation of escape dynamics can be gained by
considering the following solution

\begin{equation}\label{bsol}g(t,~z)= e^{-iE_0t}\int\limits_p\limits^{\infty}dEe^{-iEt}\varphi_p(E,~z)\left<\varphi_p(E)|g(0)\right>~,\end{equation}
where $g(0,~z)$ denotes the brane localized initial state. So that
the resulting prescription for evaluating the probability of
particle to be confined at instant $t$ on the brane
\[\left|\left<g(0)|g(t)\right>\right|^2~,\] is similar to the
quantum mechanical one considered in \cite{FK, CSMFGR}.
Correspondingly the time evolution of the decay can be divided
into three domains as in the quantum mechanical case indicated
above. Under assumption $\mu/k \ll 1$ one finds that the function
$\left|C(E)\right|^2$ where
\[C(E)=\left<\varphi_p(E)|g(0)\right>~,\]
has a simple pole in the fourth quadrant of a complex $E$ plane
\cite{Ma, DRT}
\[\tilde{E}=\sqrt{\vec{p}\,^2+m_0^2}-{im_0\Gamma\over \sqrt{\vec{p}\,^2+m_0^2}}~.\] Since, when $t > 0$ and
$|E|\rightarrow\infty$, $e^{-iEt}\rightarrow 0$ in the fourth
quadrant, for evaluating of the transition amplitude
\[\left<g(0)|g(t)\right>=e^{-iE_0t}\int\limits_p\limits^{\infty}dEe^{-iEt}\left|C(E)\right|^2~,\]
it is convenient to deform the integration contour as it is shown
in Fig.1. In this way one finds
\begin{eqnarray}\int\limits_p\limits^{\infty}dEe^{-iEt}\left|C(E)\right|^2=\mbox{residue
term}-\nonumber\\ie^{-ipt}\int\limits_0\limits^{\infty}dEe^{-Et}\left|C(p-iE)\right|^2~,\end{eqnarray}
where the residue term determines the exponential decay law with
the decay width $\sim \Gamma  m_0/\sqrt{\vec{p}\,^2+m_0^2}$. In
the vicinity of the pole $\tilde{m}=m_0-i\Gamma$
\[\varphi(m,~z)\varphi(m,~z')\sim {d(z,~z')\over m-\tilde{m}}\approx
{m\over \tilde{E}}{d(z,~z')\over E-\tilde{E}}~.\] Thus the
pre-exponential term that comes from the residue at the pole
\[-2\pi i\lim\limits_{E\rightarrow \tilde{E}}\left(E-\tilde{E}\right)\left|C(E)\right|^2~,\]
does not depend on the three momentum $\vec{p}$. By taking into
account that $\mu/k \ll 1$ and the localization width of the
scalar is $\sim k^{-1}$, one can simply take
\begin{equation}\label{initstate}g(0,~z)=\left\{\begin{array}{ll}\sqrt{k}/\sqrt{1-e^{-1}}~, & \mbox{for}~~ |z|\leq (2k)^{-1}~,\\
0~, & \mbox{for}~~ |z|>(2k)^{-1}~.
\end{array}\right.\end{equation} So using the solution Eq.(\ref{bsol}), on the quite general
grounds as in the standard quantum mechanical case \cite{CSMFGR}
one concludes that initially the decay is slower than exponential,
then comes the exponential region and after a long time it obeys a
power law. We think the Eq.(\ref{bsol}) is of particular
importance for describing the dynamics of particle escape into
extra space. In studying this problem for arbitrary spin integer
particle one can construct analogous solution.

Let us make a brief clarification of the above discussion. Usually
in constructing the relativistic quantum mechanics the square root
from the Hamiltonian of a relativistic free particle
\begin{equation}\label{srelpar}i\partial_t\psi=\sqrt{H}\psi~,\end{equation} is removed. The resulting Klein-Gordon
equation
\begin{equation}\label{KG}-\partial_t^2\psi=H\psi~,\end{equation} is
altered from the Schr\"{o}dinger equation in that it contains the
second order time derivative that in general proves impossible the
probability interpretation of the solution. Knowing the
eigenfunction spectrum \[H\varphi=E^2\varphi~,\] loosely speaking
one can construct the solution of Eq.(\ref{srelpar})
\[\psi(t)=\int dEe^{-iEt}\varphi(E)\left<\varphi(E)|\psi(0)\right>~,\] satisfying the Eq.(\ref{KG}) as well.
This solution admits a straightforward probability interpretation
as in the Schr\"{o}dinger case and is thereby unique. This is
essentially what we have done in the present paper for describing
the time evolution of the quasilocalized scalar particle on the
brane. This brief paper corrects and complements our previous
consideration \cite{Ma}.

\vspace{0.2 cm}

\centerline{\bf Acknowledgements} The author is greatly indebted
to B.~Bajc, Z.~Berezhiani, G.~Gabadadze and A.~Rosly for very
useful discussions and comments. The work was supported by
\emph{Georgian President Fellowship for Young Scientists} and the
grant FEL. REG. $980767$.

%%%%%%%%%%%%%%%%%%%%%%%%%%%%%%%%%%%%%%%%%%%%%%%%%%%%%%%%%%%%%%%%%%%%%%%%


\begin{thebibliography}{10}

\bibitem{Ma}
M.~Maziashvili, Phys. Lett. B627 (2005) 197, hep-ph/0507103.

\bibitem{BG}
B.~Bajc and G.~Gabadadze, Phys. Lett. {\bf B474} (2000) 282,
hep-th/9912232.

\bibitem{GoRa}
M.~Gogberashvili, Int. J. Mod. Phys. {\bf D11} (2002) 1635,
hep-th/9812296; M.~Gogberashvili, Int. J. Mod. Phys. {\bf D11}
(2002) 1639, hep-ph/9908347; L.~Randall and R.~Sundrum, Phys. Rev.
Lett. {\bf 83} (1999) 3370, hep-ph/9905221; L.~Randall and
R.~Sundrum, Phys. Rev. Lett. {\bf 83} (1999) 4690, hep-th/9906064.

\bibitem{DRT}
S.~Dubovsky, V.~Rubakov and P.~Tinyakov, Phys. Rev. {\bf D62}
(2000) 105011; hep-th/0006046; V.~Rubakov, Phys. Usp. {\bf 44}
(2001) 871 (Usp. Fiz. Nauk {\bf171} (2001) 913); hep-ph/0104152.

\bibitem{FK}
V. Fock and N. Krylov, Journal of Physics 11 (1947) 112.

\bibitem{CSMFGR}
C.~Chiu, E.~Sudarshan and B.~Misra, Phys. Rev. {\bf D16} (1977)
520; L.~Fonda, G.~Ghirardi and A.~Rimini, Rept. Prog. Phys. {\bf
41} (1978) 587.

\bibitem{Lee}
T.~D.~Lee, \emph{Mathematical Methods of Physics} (Moscow, Mir,
1965).



\end{thebibliography}
\end{document}